# Demonstration of magnetic dipole-dipole interaction by using smartphone pressure sensor

Sanjoy Kumar Pal[1,3], and Pradipta Panchadhyayee[2,3*]

[1]Anandapur H.S. School, Anandapur, Paschim Medinipur, West Bengal, India
[2]Department of Physics (UG & PG), Prabhat Kumar College, Contai, Purba Medinipur, India
[3]Institute of Astronomy Space and Earth Science, Kolkata -700054, W. B., India
[*]E-mail: ppcontai@gmail.com

## Abstract

In this paper, we present a hands-on activity designed to verify the dependence of the magnetic force between two identical N35 neodymium disc magnets on their separation distance. Utilizing a weight-measuring device incorporating a smartphone pressure sensor placed inside an inflated Ziplock bag, with a glass plate ensuring perfect contact, we measured the magnetic force with high precision. Our results confirm the established inverse fourth power relationship between magnetic force and distance. The linear plot of magnetic force versus the inverse fourth power of distance corroborates the corresponding theoretical model. From the slope of this linear plot, we have calculated the magnetic dipole moment of each magnet, providing a practical validation of theoretical predictions. This methodology also offers an effective approach for educational and experimental verification of magnetic interactions.

## Introduction

Smartphones are widely used in society nowadays, even among students in their day-to-day lives. Smartphones are used for entertainment and also for browsing social media. However, smartphones are powerful learning tools as well. Most smartphones are equipped with numerous sensors that make them smart devices. By utilizing different sensors like the accelerometer, gyroscope, proximity sensor, light sensor, magnetic sensor, sound sensor, camera sensor, and pressure sensor, one can conduct various science experiments [1,2]. Students can leverage smartphones as portable science laboratories, enabling them to perform a wide range of experiments with the proper use of smartphones. They can measure acceleration due to gravity, the velocity of sound, magnetic force, the spring constant of a spring, the distance of a virtual image, and many more [3-13]. Besides, the determination of the refractive index of water and glass as well as the verification of Newton's third law is also possible using smartphone sensors [14-15]. By exploring the 'smartphone lab' and applying various sensors to science experiments, students can gain a better understanding of science through hands-on activities. Students can learn science with hands-on activity more easily, which helps them to learn science theory more easily [16-17]. Here, we describe a classroom-friendly experiment using the pressure sensor of a smartphone. The pressure sensor is a common built-in sensor in smartphones. This sensor provides real-time atmospheric air pressure values through various applications like Phyphox, Physics Toolbox Sensor, and Arduino SJ available on the Play Store and App Store. In this article, we mainly use the pressure sensor of a smartphone as a force-measuring device. We take two identical thin neodymium disc magnets and fix them in 'repulsive' mode on a glass plate. The magnet system is placed on a smartphone enclosed by an airtight Ziplock bag. By varying the distance between the magnets, the readings for the magnetic forces are recorded in terms of corresponding pressure readings using the smartphone's pressure sensor. By analyzing the pressure data, we verify that the dependence of magnetic force between the two disc magnets on their internal distance conforms with the reported variation in the case of similar systems. We have also calculated the magnetic dipole moment of the magnets.

**Theoretical background**

The force acting between two identical disc magnets with the vectors of magnetization on their common axis is approximate by the expression [18,19]

$$F(x) = \frac{\pi\mu_0 M^2 R^4}{4}\left[\frac{1}{x^2} + \frac{1}{(x+2d)^2} - \frac{1}{(x+d)^2}\right], \quad (1)$$

where $\mu_0$ is the magnetic permeability in free space, and $M$ is the magnetization of the magnets. Radius and thickness of each of both identical magnets are $R$ and $d$, respectively, whereas $x$ is the distance between the magnets.
If $V$ is the volume of each magnet, the effective magnetic dipole moment of each magnet, $m = MV$.
As the magnets are cylindrical, so $V = \pi R^2 d$
When $d \ll x$ the point dipole approximation is obtained, Eq.(1) reduces to

$$F(x) = \frac{3\pi\mu_0}{2} M^2 R^4 d^2 \frac{1}{x^4} = \frac{3\mu_0}{2\pi} M^2 V^2 \frac{1}{x^4} = \frac{3\mu_0}{2\pi} m^2 \frac{1}{x^4} \quad (2)$$

The equation (2) highlights that the magnetic force follows an inverse fourth-power relationship with distance. In our experiment, we focus on demonstrating the nature of variation magnetic dipole-dipole interaction with the distance between them using smartphone pressure sensors.

**Experimental**

In view of the above discussion, we have used two thin identical neodymium disc magnets (30 mm in diameter and 5 mm in thickness) and measured the magnetic force between the magnets with varying distances. Our aim is to validate the linear relationship between $F(x)$ and $x^{-4}$ by experimental observation.

Initially, we activate the pressure sensor in the Phyphox application on an iPhone 12 Pro Max. A transparent, high-quality Ziplock bag is taken in the inflated condition. The bag is inflated with sufficient air to make its volume nearly maximum. At this condition, the bag is closed with its lock and becomes airtight. Next, we have selected a rectangular glass plate whose area is less than the flat area of the top side of the inflated Ziplock bag. We have placed a glass plate on top of the Ziplock bag. Such selection of the glass plate on the basis of its area ensures that the whole bottom area of the glass plate is in perfect contact with the Ziplock bag. The total system acts as a weight-measuring device and becomes ready for use. However, confusion may arise due to the flexibility of the Ziplock bag, which can cause a little change in the effective area corresponding to the applied force. To fix this issue, we have recorded pressure readings for various known weights centrally positioned on the glass plate (see Fig 1a). These readings are plotted in Fig. 1b which provides a slope of 0.1714 kg/hPa. Consequently, we have determined the effective area ($\propto$) as 0.0168 m².

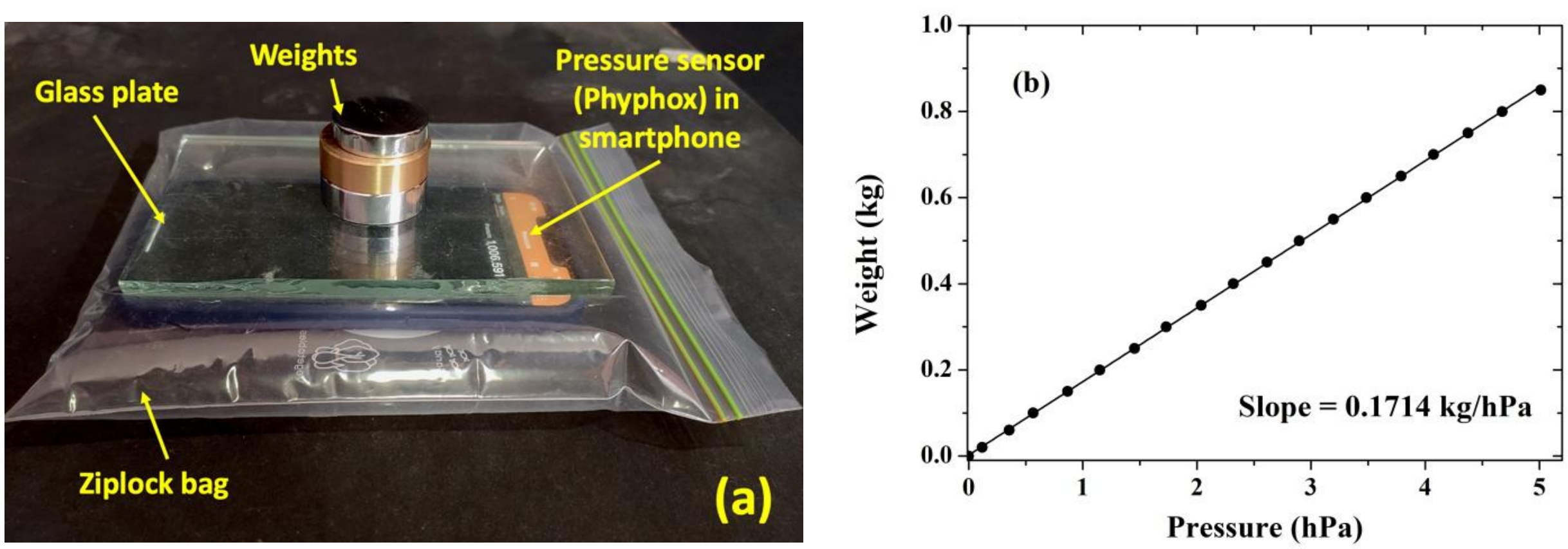

Fig. 1: (a) Weights are placed on a glass plate and a smartphone (Phyphox pressure sensor on) is inside the Ziplock bag to calibrate the effective area. (b) Weight vs pressure graph.

Table 1: Table for plotting the distance versus reading
Initial pressure, $P_0$=1007.064 hPa; Effective area, $\propto$= 0.0168 m²

| Distance, $x$ (m) | Pressure $P_i$ (hPa) | Pressure $P = P_i - P_0$ (hPa) | Force $F(x)$ (N) |
|---|---|---|---|
| 0.062 | 1007.093 | 0.029 | 0.0487 |
| 0.057 | 1007.122 | 0.058 | 0.0974 |
| 0.049 | 1007.181 | 0.117 | 0.1966 |
| 0.040 | 1007.350 | 0.286 | 0.4805 |
| 0.030 | 1007.660 | 0.596 | 1.0013 |
| 0.027 | 1008.103 | 1.039 | 1.7456 |
| 0.025 | 1008.372 | 1.308 | 2.1976 |
| 0.023 | 1008.944 | 1.880 | 3.1586 |
| 0.020 | 1010.246 | 3.182 | 5.3461 |
| 0.019 | 1011.297 | 4.233 | 7.1119 |

To set up the next part of the experiment (Fig. 2a), we safely have clamped a PVC pipe vertically using a retort clamp with a stable stand. A neodymium magnet (A) is affixed to the lower side of the PVC pipe using a double-sided adhesive tape. It is crucial to set the position of the A magnet and the pipe accurately. We have now placed the smartphone (iPhone 12 Pro Max) into the central position of the inflated Ziplock bag in air-tight condition. The glass plate under consideration is placed centrally on the Ziplock bag. On this glass plate, we have attached another identical magnet B using the double-sided adhesive tape. Initially, we take the pressure reading $P_0$ on the phyphox pressure sensor, when the magnet B fixed on the glass plate with the bag-smartphone system as a measuring arrangement. The A magnet, fixed on the PVC pipe, is placed vertically above the B magnet in 'repulsive' mode by a long distance ($x$) from the magnet B. A plastic scale of length 30 cm is employed to measure the distance between the two magnets accurately (see Fig 1a). The scale is erected and held vertically. It is attached with the PVC pipe in the parallel direction using an adhesive tape. The first distance should be tuned properly so that the $x$ value can be decreased to have number of observations (minimum 10) maintaining a regular interval. In this orientation, the pressure reading is recorded by the pressure sensor of the smartphone ($P_i$). The distance between the magnets is then gradually decreased, as stated before, and also the corresponding pressure reading is noted (see Table 1). By multiplying pressure difference, $P = P_i - P_0$, by effective area ($\alpha$) the magnetic force ($F(x)$) at a distance ($x$) acting on the glass plate is measured. The computed data of $F(x)$ are also tabulated in the last column of Table 1. As the internal distance between the magnets decreases, the magnitudes of magnetic force are found to increase.

We have plotted a graph of $x$ versus $F(x)$, which is best fitted with the relation, $F(x) = 1.15974 \times 10^{-6} x^{-4}$. The exponent of the reciprocal of $x$ is calculated as 3.928. This value clearly indicates that magnetic force varies inversely proportional to the fourth power of internal distance between the magnets (see Fig. 2b). Thus, our observation supports the nature of variation of magnetic dipole-dipole interaction with distance [Eq. 2]. The obtained value of exponent of the reciprocal of distance differs from the reported value by only 1.8%. Additionally, we have plotted a graph to show the variation of the magnetic force with the fourth power of the reciprocal of distance (Fig. 2c), which is found exhibit the expected linear relationship. The slope, denoted as $k = \frac{3\mu_0 m^2}{2\pi}$, is computed as $8.95 \times 10^{-6}$ $Nm^4$. Here, $m$ in the expression is the magnetic dipole of each of the disc magnets. From the value of the slope, the value of the magnetic dipole moment is determined as 1.22 $Am^2$.

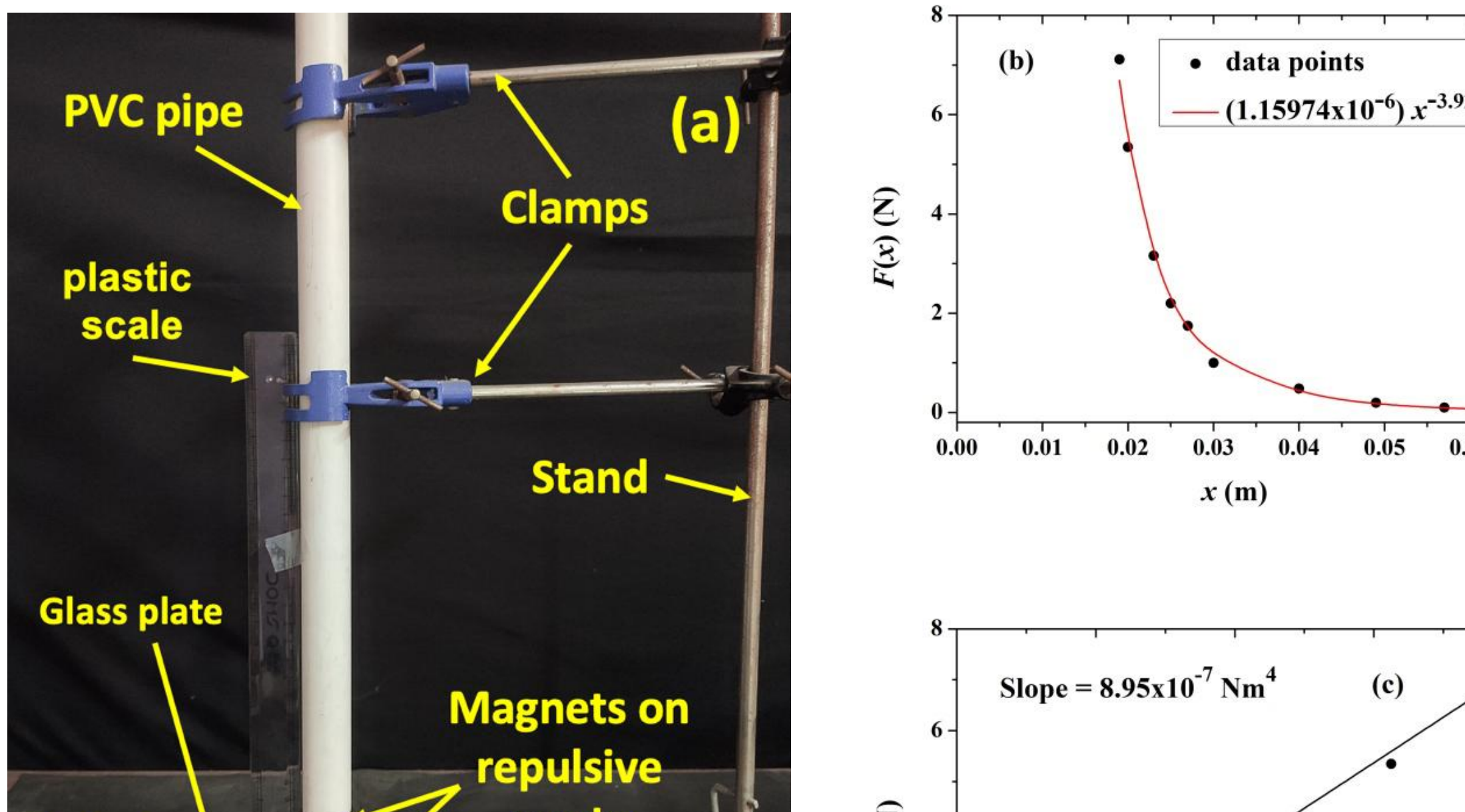

Fig. 2: (a) Experimental setup; (b) Magnetic force versus distance graph; (c) Magnetic force versus $(\text{distance})^{-4}$ graph

**Conclusion**

In this paper, we demonstrate a hands-on activity to verify the nature of magnetic force. Using smartphone pressure sensor we measure the magnetic force between two neodymium magnets. For the measurement of force we have used a glass plate located on the Ziplock bag on perfect touch condition. We have measured the effective touch area of the glass plate and Ziplock bag. After that we observes the variation of distance vs force. We have plotted the graph between force vs distance and it shows a clear inverse forth power variation of force with distance. Lastly we have plotted another graph for the variation of magnetic force with the fourth power of the reciprocal of distance with same data and get a straight-line graph with a positive slope. From the slope we have also calculated the value of magnetic dipole moment of the used magnet. Both the graph verifies the magnetic dipole-dipole interaction laws.